\documentclass{article}

\usepackage{amsmath}
\usepackage{graphicx}

\newcommand{\be}{\begin{equation}}
\newcommand{\ee}{\end{equation}}
\newcommand{\bea}{\begin{equation}\begin{aligned}}
\newcommand{\eea}{\end{aligned}\end{equation}}
\newcommand{\dd}{\partial}

\newcommand{\inst}{\mathop{\rm (inst)}\nolimits}
\newcommand{\Tr}{\mathop{\rm Tr}\nolimits}
\newcommand{\Pexp}{\mathop{\rm Pexp}\nolimits}
\newcommand{\E}{\mathop{\rm E}\nolimits}
\newcommand{\B}{\mathop{\rm B}\nolimits}
\newcommand{\sh}{\mathop{\rm sh}\nolimits}
\newcommand{\ch}{\mathop{\rm ch}\nolimits}

\title{Non-Abelian field strength bilocal correlator at finite temperature
in the model of dilute instanton gas}
\author{N.O.~Agasian \thanks{agasian@heron.itep.ru} \and S.M.~Fedorov \thanks{fedorov@heron.itep.ru} }
\date{June 5, 2004}

\begin{document}
\unitlength=1cm

\maketitle
\vspace{-1cm}
\begin{center}
{\it Institute of Theoretical and Experimental Physics, \\117218, Moscow,
B.Cheremushkinskaya 25, Russia}
\end{center}
\vspace{1cm}

\begin{abstract}
Bilocal correlator in gluodynamics is evaluated at finite temperature in the framework of
instanton gas model. It is demonstrated that vacuum correlation length decreases
with a growing temperature. Obtained results are compared with lattice data for
the bilocal correlator at finite temperature. Density of instantons and possible
structures of nonperturbative vacuum are discussed.
\end{abstract}


\section{Introduction}
Nonperturbative fields play important role in the vacuum of QCD and gluodynamics,
and determine many properties of the theory, in particular,
they are responsible for the confinement, spontaneous chiral symmetry breaking (SCSB),
and as a result, for the formation of observed hadron spectrum.
A well developed and successful in the description of many properties of QCD
model is the instanton liquid model, suggested by
Shuryak~\cite{Shuryak_81} and Diakonov and Petrov~\cite{Diak_Pet_84}.
The main nonperturbative fields in this picture are well separated instantons
and antiinstantons, their interaction being not very strong (thus the name ''liquid'').
Instanton density is approximately $1$~fm$^{-4}$. This model solves a number of problems
in QCD, in particular it naturally leads to SCSB, explains the $\eta'$ meson mass.
However, there is a number of problems of fundamental importance:
it remains unclear, how to make such instanton-antiinstanton ensemble
stable (infrared inflation of instantons), and confinement cannot be explained in the
instanton liquid model.

However, apart from quasiclassical instantons there are other nonperturbative fields in
the vacuum, which allow to solve "infrared problem" of instantons.  Nonperturbative
vacuum of QCD can be  parametrized by gauge invariant vacuum averages of gluon
field strength tensor -- vacuum correlators~\cite{Dosch_87}. This allows to
successfully describe many phenomena in QCD (see the review papers~\cite{DShS, SSDP}).
Bilocal correlator is sufficient for qualitative description of many phenomena
(stochastic vacuum model). Moreover, there are indications that corrections due
to higher correlators are small and in some cases amount to several percent~\cite{DShS}.
Nonzero string tension naturally appears in this approach and can be expressed through
vacuum average of chromoelectric field~\cite{Dosch_87}
\be
\label{eq_sigma}
\sigma = \frac{g^2}{2} \int d^2 x \langle \Tr\left( E_i(x) \Phi(x,0) E_i(0) \Phi^{\dagger}(x,0) \right)\rangle,
\ee
where $E_i=F_{4i}$ is chromoelectric field and $F_{\mu\nu} = F_{\mu\nu}^a t^a$ is the field strength tensor of
gluon field. $\Phi(x,0)$ is the parallel transporter along the straight line,
connecting points $x$ and $0$. Integration in~(\ref{eq_sigma}) goes in space-time plane ($i$,4).

The problem of infrared inflation of instantons is also naturally solved in the stochastic vacuum model.
Investigation of influence of nonperturbative quantum fluctuations on instantons was started in
papers~\cite{SVZ,MAK}. Later it was shown that standard perturbation theory changes in the
nonperturbative stochastic vacuum, and contribution of large instantons to physical quantities
becomes finite~\cite{Agasian:1994yp}. Next, direct interaction of instanton with nonperturbative vacuum
fields was shown~\cite{Agasian:2001vw} to lead to stabilization in instanton size, the average size of instantons
being of the order of vacuum correlation length. Distribution of instantons over sizes was found to be
in good agreement with lattice data, with a maximum at $\rho_c \simeq 0.25\div 0.3$~fm.

An important property of vacuum is the scaling of string tensions between sources in different representations
of color group, which is observed in the lattice calculations~\cite{Bali:2000un}. It was
shown in~\cite{Shevchenko:2000du} that instanton contribution to the force between
heavy quarks in different representations violates Casimir scaling, and this
leads to a constraint on instanton density.

In the present paper we study instantons' contribution to the bilocal correlator
at finite temperature\footnote{
Bilocal correlator in the instanton gas at zero temperature
was evaluated in~\cite{Dorokhov:1997iv,Ilgenfritz:1997cd}.
}.
We demonstrate that correlation length of bilocal correlator,
evaluated on calorons (calorons are instanton-like configurations at finite temperature) in the lowest
order in caloron density has significant dependence on temperature. At the same time
it is found in the lattice calculations that within errors correlation length for
chromomagnetic fields does not depend on the temperature in the whole region between
zero and critical temperature $T_c$~\cite{D'Elia:2002ck}.

Temperature dependencies of the chromomagnetic bilocal correlator and of the spatial string tension,
$\sigma_s(T)$, were found analytically
in the paper~\cite{Agasian:2003yw}. It was shown there that chromomagnetic condensate
slowly grows with increasing temperature up to $2 T_c$, while correlation length $\lambda_m(T)$
does not depend on $T$ in this temperature region.
At high temperature, $T>2 T_c$, condensate grows with $T$ as $\langle H^2 \rangle_T \propto g^8(T) T^4$
and magnetic correlation length decreases as $\lambda_m(T) \propto 1/(g^2(T) T)$.
Obtained analytical expression for the $\sigma_s(T)$ agrees well with lattice results
for all temperatures~\cite{Boyd:1996bx}.
There exists a simple physical explanation of the independence of correlation length on
the temperature at $T\le T_c$. Inverse correlation length $M=1/\lambda_m$
is a mass of lightest gluelump $1^{+-}$.
Gluelumps~\cite{Campbell:1985kp} are not physical objects, however they play
an important role in the nonperturbative QCD, because their masses determine
field correlators, and in particular, string tension at $T=0$ is related to gluelumps.
Their masses were calculated analytically in the framework of QCD sum rules~\cite{Dosch:1998th},
QCD string model~\cite{Simonov:2000ky} and were determined numerically in various
lattice calculations~\cite{Philipsen:2001ip}. Summing up all those results,
correlation length equals $\lambda_m \approx 0.15$~fm, which corresponds to
$M\approx 1.5$~GeV. Taking into account that phase transition temperature
is much less than gluelump mass $T_c \ll M$, it is physically clear that
temperature dependence of gluelump mass, and therefore of the correlation length,
should be very weak in the low temperature region $T<T_c \simeq 300$~MeV,
which is confirmed by analytical~\cite{Agasian:2003yw} and lattice~\cite{D'Elia:2002ck}
calculations.
It should be mentioned, that magnetic correlation length does not change it's behavior
at the phase transition point. It is well known, that order parameter for the
deconfinement phase transition is vacuum average of Polyakov loop, and therefore
it is Polyakov loops correlator $\langle L(\vec x) L^{\dagger}(0) \rangle$
that significantly changes it's behavior at temperature $T=T_c$.

\section{Chromoelectric and chromomagnetic correlators}

At finite temperature euclidian O(4) space-time symmetry is broken to space O(3) symmetry,
and bilocal correlator is described by independent chromomagnetic and chromoelectric
correlation functions, which are expressed through four independent
functions $D^{\E}(x^2)$, $D_1^{\E}(x^2)$, $D^{\B}(x^2)$ ¨ $D_1^{\B}(x^2)$:
\bea
&g^2\langle \Tr\left( E_i(x) \Phi(x,y) E_j(y) \Phi^{\dagger}(x,y) \right)\rangle =
  \delta_{i j} \left( D^{\E} + D_1^{\E} + z_4^2 \frac{\dd D_1^{\E}}{\dd z^2} \right) +
  z_i z_j \frac{\dd D_1^{\E}}{\dd z^2},\\
&g^2\langle \Tr\left( B_i(x) \Phi(x,y) B_j(y) \Phi^{\dagger}(x,y) \right)\rangle =
  \delta_{i j} \left( D^{\B} + D_1^{\B} + \vec{z}^2 \frac{\dd D_1^{\B}}{\dd z^2} \right) -
  z_i z_j \frac{\dd D_1^{\B}}{\dd z^2},
\eea
where $B_i =\frac12 \varepsilon_{ijk} F_{jk}$ is the chromomagnetic field,
$E_i=F_{4i}$ -- chromoelectric field and $F_{\mu\nu} = F_{\mu\nu}^a t^a$ is the
field strength tensor.
\be
\Phi(x,y) = \Pexp \left( i g z^{\mu} \int_0^1 ds \, A_{\mu}(y + sz) \right)
\ee
is the parallel transporter along the straight line, connecting points $x$ and $y$; $z = x-y$.
These phase factors $\Phi$  are introduced to ensure gauge invariance of
correlators.
Obviously $D_1^{\E} = D_1^{\B} = 0$, $D^{\E} = D^{\B}$ for the
selfdual field (see~\cite{Ilgenfritz:1997cd}).
Thus it is sufficient to consider chromomagnetic correlator only.

The very fact that for instantons $D^{\E} = D^{\B}$ implies that instantons cannot be the only
nonperturbative fields in the vacuum of gluodynamics, because it is known from lattice
calculations that chromoelectric and chromomagnetic condensates have different dependences
on temperature. In particular, at the deconfinement phase transition temperature chromoelectric
condensate abruptly vanishes, while chromomagnetic condensate stays intact~\cite{D'Elia:2002ck}.

\section{Calorons' contribution to the chromomagnetic correlator}

Instanton field with the center at the origin in the singular gauge is given by
\bea
A_{\mu}^{{a \inst}}(x) = -\frac{1}{g} \bar{\eta}_{\mu\nu}^a \dd_{\nu} \ln \Pi^{\inst}(x), \\
\Pi^{\inst}(x) = 1 + \frac{\rho^2}{x^2} = 1 + \frac{\rho^2}{r^2+\tau^2},
\eea
where $x = (\vec{r},\tau)$. This expression is easily generalized to the case
of finite temperature, and caloron has a well known form~\cite{Harrington:1978ve}:
\bea
&A_{\mu}^{a}(x) = -\frac{1}{g} \bar{\eta}_{\mu\nu}^a \dd_{\nu} \ln \Pi(x), \\
&\Pi(x) = 1 + \sum_{n=-\infty}^{\infty} \frac{\rho^2}{r^2+(\tau-n\beta)^2}
       = 1 + \frac{\pi \rho^2}{\beta r} \frac{\sh(2 \pi r/\beta)}{\ch(2\pi r/\beta) - \cos(2 \pi \tau/\beta)},
\eea
where $\beta=1/T$ is the inverse temperature.
Caloron field is given by
\bea
&F_{\mu\nu}^a = \dd_{\mu} A_{\nu}^a - \dd_{\nu}A_{\mu}^a + g \varepsilon^{a b c}A_{\mu}^b A_{\nu}^c,\\
&g B_i^a = -\delta_{a i} g_1(x) + \varepsilon_{a i j} \frac{x_j}{r} g_2(x) + \frac{x_a x_i}{r^2} g_3(x),\\
&E_i^a = B_i^a,\\
&g_1(x) = \frac{1}{r}\frac{1}{\Pi} \frac{\dd \Pi}{\dd r} +
          \frac{1}{\Pi}\frac{\dd^2 \Pi}{\dd \tau^2} -
          \frac{1}{\Pi^2} \left( \frac{\dd \Pi}{\dd\tau} \right)^2 +
          \frac{1}{\Pi^2} \left( \frac{\dd \Pi}{\dd r} \right)^2,\\
&g_2(x) = -\frac{2}{\Pi^2} \frac{\dd \Pi}{\dd r}\frac{\dd \Pi}{\dd \tau} +
          \frac{1}{\Pi}\frac{\dd^2 \Pi}{\dd r \dd \tau},\\
&g_3(x) = \frac{2}{\Pi^2} \left( \frac{\dd \Pi}{\dd r}\right)^2 -
          \frac{1}{\Pi}\frac{\dd^2\Pi}{\dd r^2} +
          \frac{1}{r} \frac{1}{\Pi} \frac{\dd \Pi}{\dd r}.
\eea
Owing to the selfduality of caloron field $F_{\mu\nu} = \tilde F_{\mu\nu}$
chromoelectric field is equal to chromomagnetic one.

Next step is the evaluation of  the parallel transporter on caloron:
\bea
  &\begin{aligned}
  \Phi(x,y) &= \Pexp \left( i g z^{\mu} \int_0^1 ds \, A_{\mu}(y + sz) \right) =
    \Pexp \left( -i z^{\mu} \bar{\eta}_{\mu\nu}^a t^a \int_0^1 ds \, \dd_{\nu} \ln \Pi (y+sz) \right) = \\
  &= \Pexp \left( -i t^a \int_0^1 ds \left[
    \bar{\eta}_{i j}^a z_i \dd_j \ln\Pi + \bar{\eta}_{i 4}^a z_i \dd_4 \ln\Pi +
    \bar{\eta}_{4 j}^a z_4 \dd_j \ln\Pi
    \right] \right),
  \end{aligned} \\
&z = x-y.
\eea

For comparison with lattice results we will consider the case $z_4 = 0$.
Then
\bea
\Phi(x,y) &= \Pexp \left( -i t^a \int_0^1 ds \left[
  \bar{\eta}_{i j}^a z_i \dd_j \ln\Pi + \bar{\eta}_{i 4}^a z_i \dd_4 \ln\Pi
  \right] \right).
\eea

In general one has to evaluate path ordered integral. However there are
special cases, when the exponent commutes at different values of $s$. Obviously,
these cases are
$\tau = 0$, $\tau = \beta/2$ and $\tau = \beta$, since $\dd_4 \ln \Pi (r,\tau = 0;\beta/2;\beta) = 0$.
Another example is the single instanton field, since $\dd_4 \ln\Pi^{\inst} = \tau (1/r) (\dd \ln\Pi^{\inst}/\dd r)$.

To evaluate the contribution of caloron gas to the two-point correlator in the
lowest order in density one has to average single-caloron contribution over it's position,
or, which is the same, to average over $y$ at a given $z=x-y$.

Averaging over $y$ comes to the integral $\int d^4 y = \int d^3 \vec{y} \int_0^{\beta} d y_4$.
Averaging over three-dimensional vector $\vec{y}$ does not make any problems, but the integral over $y_4$
is not tractable, since correlator can be numerically calculated only when $\tau = 0, \beta/2, \beta$.
Nevertheless it will be seen from comparison with the single-instanton case that at low temperatures
calculations with $\tau=0$ are sufficient, and at high temperatures correlation lengths for
$\tau = 0$ and $\tau = \beta/2$ cases coincide, which makes averaging trivial.

\section{Numerical results}

Let us use the equation
\be
g^2\langle \Tr\left( B_i(x) \Phi(x,y) B_j(y) \Phi^{\dagger}(x,y) \right)\rangle_{\vec{y}} =
  \delta_{i j} \left( d + d_1 + \vec{z}^2 \frac{\dd d_1}{\dd z^2} \right) -
  z_i z_j \frac{\dd d_1}{\dd z^2},
\ee
at $z_4 = 0$ to define the functions $d$ and $d_1$, which depend on $z^2$ and $y_4 \equiv \tau$:
$d = d(\tau,z^2); d_1 = d_1(\tau,z^2)$. Averaging over $y_4$ yields
\bea
&\int_0^{\beta} d \tau d(\tau, z^2) = D^{\B}(z^2),\\
&\int_0^{\beta} d \tau d_1(\tau, z^2) = D_1^{\B}(z^2) = 0,\\
&D^{\B}(z^2=0) = \frac{ 4\pi^2}{3}.
\eea
Last equality follows from the relation
$g^2 \int d^3 \vec{y} \int_0^{\beta} d y_4 \left( F_{\mu\nu}^a\right)^2 = 32 \pi^2$
(the field is selfdual, and this integral is proportional to the topological charge $\sim \int d^4 y F \tilde F$).

Let us define correlation length $\lambda_{\tau}$ as the coefficient in the exponent of the best fit to the
function $d(\tau,z^2)$:
\be
d(\tau, z^2) \simeq e^{-|z|/\lambda_{\tau}}
\ee

\begin{figure}[htb]
\begin{picture}(15,6.2)
\put(0.7,0.6){\includegraphics[height=5.5cm]{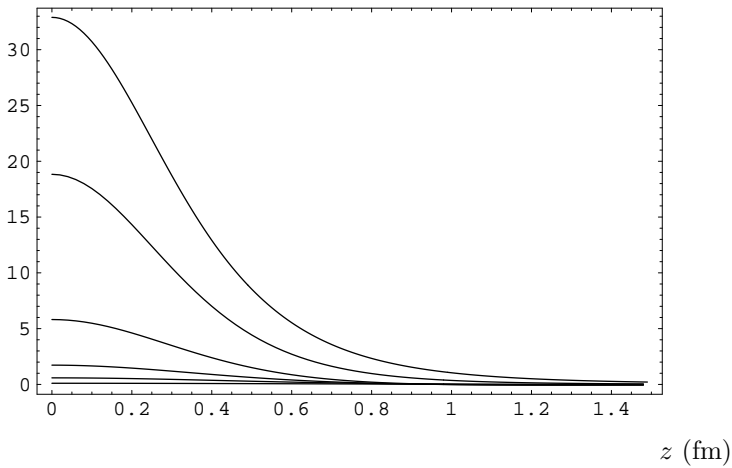}}
\put(9.5,0.1){$z$ (fm)}
\end{picture}
\caption{
$d^{\inst}(\tau,z^2)$ at $\tau=(0, 0.15, 0.3, 0.45, 0.6)$~fm; $\rho=0.3$~fm. Upper line
corresponds to $\tau=0$.
}
\label{fig_inst_dz}
\end{figure}

\begin{figure}[htb]
\begin{picture}(15,6.2)
\put(0.7,0.6){\includegraphics[height=5.5cm]{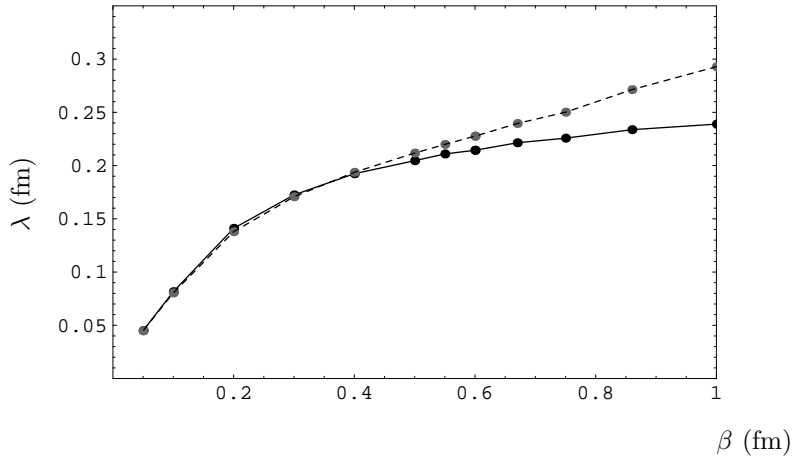}}
\put(9.5,0.1){$\beta$ (fm)}
\put(.1,3){\rotatebox{90}{$\lambda$ (fm)}}
\end{picture}
\caption{Correlation length of functions $d(\tau=0,z^2)$ and $d(\tau=\beta/2,z^2)$:
$\lambda_{\tau=0}(\beta)$ (solid line) and $\lambda_{\tau=\beta/2}(\beta)$ (dashed line); $\rho=0.3$~fm.
\label{fig_lambda2_t}
}
\end{figure}

As discussed above, we will evaluate $d(\tau=0, z^2)$ and $d(\tau=\beta/2, z^2)$ only.
However, this information is sufficient to determine the correlation length of the function $D^{\B}(z^2)$.
Indeed, by averaging over $\tau$ one can find that normalized to $1$ functions
$d(\tau=0,z^2)/d(\tau=0,z^2=0)$ and $D^{\B}(z^2)/D^{\B}(z^2=0)$  practically coincide.
The fact is that instanton is a well localized field configuration, and it's field
decreases very fast when moving away from it's center. This is illustrated in the Fig.~\ref{fig_inst_dz}.
It shows function $d^{\inst}(\tau,z^2)$, calculated for the single instanton configuration for several
values of $\tau$. Profile of the function $D^{\B}(z^2)$ is determined by the function $d(\tau=0,z^2)$.
The same is true for caloron when $\beta \gg \rho$, because in this case instantons forming caloron
are well separated. At higher temperatures, when $\beta \le 2 \rho$ the functions $d(\tau=0,z^2)$ and
$d(\tau=\beta/2,z^2)$ have almost the same correlation length (see Fig.~\ref{fig_lambda2_t}),
which makes averaging over $\tau$ trivial (it only affects the amplitude of the functions, which can
be found from the condition $D^{\B}(z^2=0) = \frac{ 4\pi^2}{3}$).

Taking into account all aforesaid, we can restrict our analysis by the function $d(\tau=0,z^2)$ only.
It's correlation length is close to the correlation length of function $D^{\B}(z^2)$
for all values of $\beta$.

\begin{figure}[htb]
\begin{picture}(15,6.2)
\put(0.7,0.6){\includegraphics[height=5.5cm]{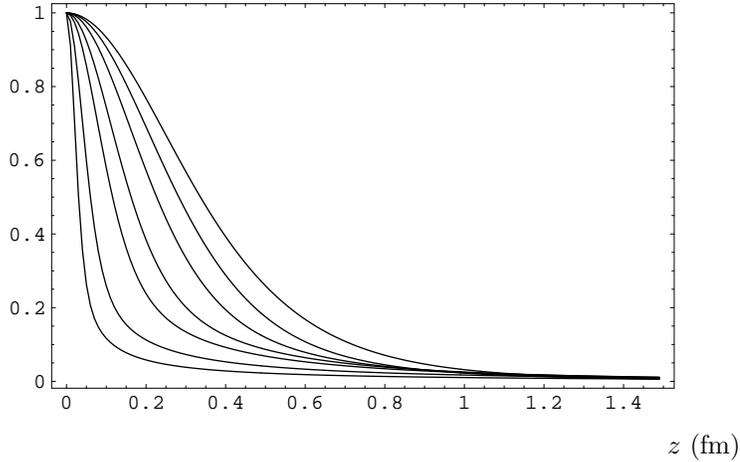}}
\put(9.5,0.1){$z$ (fm)}
\end{picture}
\caption{
$d(\tau=0,z^2)/d(\tau=0,z^2=0)$ for $\beta = (0.05, 0.1, 0.2, 0.3, 0.5, 0.86, \infty )$~fm$^{-1}$; $\rho=0.3$~fm.
The largest correlation length (upper line) corresponds to zero temperature ($\beta=\infty$),
i.e. to instantons' contribution.
}
\label{fig_corrltr}
\end{figure}

\begin{figure}[htb]
\begin{picture}(15,6.2)
\put(0.7,0.6){\includegraphics[height=5.5cm]{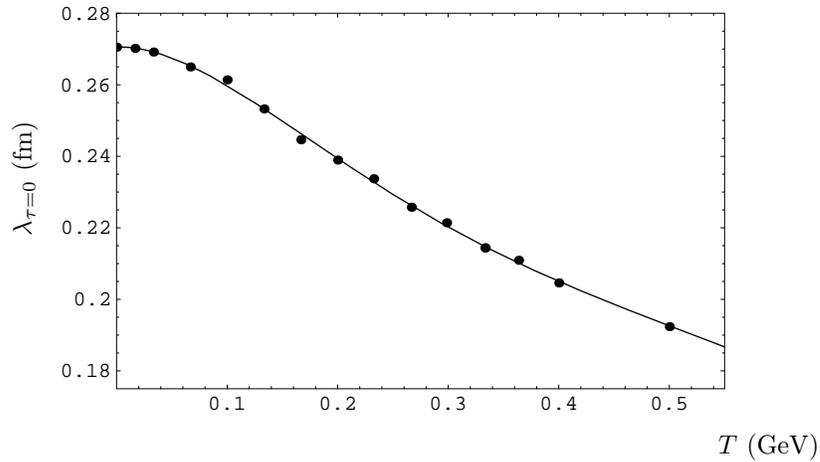}}
\put(9.5,0.1){$T$ (GeV)}
\put(.1,3){\rotatebox{90}{$\lambda_{\tau=0}$ (fm)}}
\end{picture}
\caption{Correlation length as a function of temperature $\lambda_{\tau=0}(T)$; $\rho=0.3$~fm.
}
\label{fig_lambda_t}
\end{figure}

Chromomagnetic correlator for $\tau = 0, \beta/2, \beta$ takes the form
\bea
&g^2\langle \Tr\left( B_i(x) \Phi(x,y) B_j(y) \Phi^{\dagger}(x,y) \right)\rangle_{\vec{y}} =
g^2 \int d^3 \vec{y} \left( B_i^a(x) U^{a b}(x,y) B_j^b(y) \right),\\
&U^{a b}(x,y) = \Tr \left( t_a \Phi(x,y) t_b \Phi^{\dagger}(x,y) \right) =
\frac{1}{2} \delta^{a b} \cos(2 \phi) - \frac{1}{2} \varepsilon^{a b c} n_c \sin(2 \phi) +
n_a n_b \sin^2(\phi),\\
&n_a \phi = \frac{1}{2} \bar{\eta}_{i j}^a z_i y_j \int_0^1 ds \,
  \left( \frac{1}{r} \frac{\dd \ln \Pi (y + sz)}{\dd r} \right)
\label{eq_corrlt_final}
\eea
Correlator defined in~(\ref{eq_corrlt_final}) is presented in Fig.~\ref{fig_corrltr} for several
values of $\beta$. Corresponding correlation lengths are shown in Fig.~\ref{fig_lambda_t}.

\section{Discussion}
Bilocal correlator at finite temperature has been studied on lattice in the paper~\cite{D'Elia:2002ck},
and it was found that bilocal chromomagnetic correlator practically does not depend on
temperature in the whole region from zero to $T_c$ (this dependence is also very weak above critical temperature).
At the same time, as it is seen from Fig.~\ref{fig_lambda_t}, the correlation length in the dilute instanton gas
changes by approximately $20\%$ when temperature changes from zero to $300$~MeV. Thus, one comes
to the conclusion that instanton gas cannot be a correct description of the real vacuum of gluodynamics.
There are several possible explanations of these lattice data.

\noindent
1) Instanton density is much less than $1$~fm$^{-4}$, and the main contribution
to the correlator comes from other nonperturbative fields, which are responsible for
the confinement.

\noindent
2) Instanton density is high enough to make next terms in density expansion change
the result qualitatively, leading to the independent of temperature correlation length.

\noindent
3) Instanton ensemble is rearranged at nonzero temperature, instantons and antiinstantons
form molecules, and the contribution of these instanton-antiinstanton molecules to bilocal correlator
is significantly different from that of instanton gas. This binding may be connected to the
deconfinement phase transition. For example, this is the case for 3-d adjoint Higgs model, where it was
found that in the deconfined phase instantons and antiinstantons are bound into
molecules~\cite{Agasian:1997wv}.
Formation of instanton-antiinstanton molecules in QCD vacuum and it's connection to the
chiral symmetry restoration phase transition is discussed in review~\cite{Schafer:1996wv}.

\noindent 4)
It should be noted, that in the last years a different scenario of the caloron vacuum in the
confined phase ($T < T_c$) is discussed. There exist field configurations with nontrivial
holonomy, the so called KvBLL solution~\cite{Kraan:1998sn}. This scenario~\cite{Ilgenfritz:2002qs}
suggests that calorons may be composites of dyons and antidyons. Lattice calculations, giving
numerical evidence for the existence of these field configurations for the cases
of $SU(2)$ and $SU(3)$ groups, are presented in papers~\cite{Ilgenfritz:2002yf,Gattringer:2003uq}.
Clearly, in the case of this scenario correlation length in the nonperturbative vacuum has
a more complicated temperature dependence, than in the case of caloron gas studied in this paper.

In principle, one can consider one possibility more, when instanton size distribution depends on temperature
in such a way, that the resulting correlation length does not depend on $T$. However, this is
ruled out by existing lattice calculations, which demonstrate that within errors instanton size
distribution stays the same in the temperature region from $0$ to $T_c$~\cite{Lucini:2004yh}.

In this paper we have found the contribution of instanton gas to the bilocal correlator at
finite temperature. Comparison of the temperature dependence of correlation length with lattice
results allows to make the conclusion that if instanton density is not too high and corrections to
the dilute gas approximation are not significant, then instanton density should be much less
than $1$~fm$^{-4}$, the value standard for instanton liquid model.
Vacuum of non-Abelian gauge
theory at finite temperature is a very complicated system, and it is natural to expect that it
can be a combination of the above listed possibilities.

Authors are grateful to Yu.A.~Simonov for
useful comments. This work is supported by NSh-1774.2003.2  grant and by
Federal Program of the Russian Ministry of Industry, Science
and Technology No 40.052.1.1.1112.

\end{document}